\documentclass[12pt]{article}

\usepackage{amssymb,commath,bbold}
\usepackage{color}
\usepackage{epsfig,amssymb,amsfonts,amsmath,graphicx,dsfont,cite,xfrac}
\usepackage{authblk}
\usepackage{braket,bbm}
\definecolor{mygray}{gray}{0.5}
\usepackage{subfigure}

\usepackage{cite}
\usepackage[colorlinks=true,linkcolor=blue,citecolor=red]{hyperref}

\newcommand{\be}{\begin{equation}}
\newcommand{\ee}{\end{equation}}
\newcommand{\bea}{\begin{eqnarray}}
\newcommand{\eea}{\end{eqnarray}}

\title{Optimized Binomial Quantum States of Complex Oscillators with Real Spectrum}

\author[]{Kevin D. Zelaya}
\author[]{Oscar Rosas-Ortiz}

\affil[]{\footnotesize Physics Department, Cinvestav, AP 14-740, 07000
M\'exico DF, Mexico}

\date{}
\begin{document}

\maketitle

\begin{abstract}
Classical and nonclassical states of quantum complex oscillators with real spectrum are presented. 
Such states are bi-orthonormal superpositions of $n+1$ energy eigenvectors of the system with binomial-like coefficients. For large values of $n$ these optimized binomial states behave as photon added coherent states when the imaginary part of the potential is cancelled.
\end{abstract}

\section{Introduction}

\begin{figure}[thb]
\centering
\includegraphics[width=0.3\textwidth]{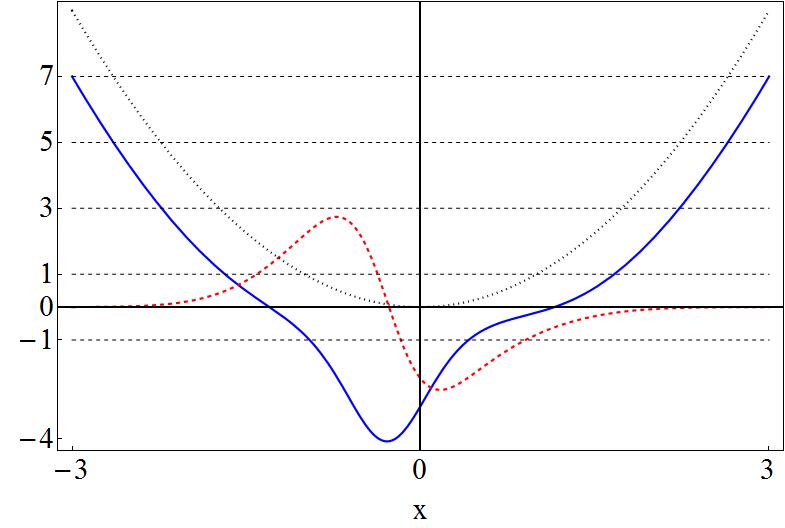}
\caption{\footnotesize 
Real (blue, continuous curve) and imaginary (red, dashed curve) parts of the complex oscillator potential $V_{\lambda}(x)$ defined in (\ref{pot1}) for $a=\tfrac{\pi}{4}$, $b= \tfrac{\sqrt{\pi}}{2}$ and $c=1$. The energy eigenvalues $E_n= 2n-1$, $n=0,1, \ldots$, are represented by dotted horizontal lines and the conventional oscillator (black, dotted curve), with eigenvalues ${\cal E}_n = 2n+1 =E_{n+1}$, has been included as a reference.}
\label{oscc}
\end{figure}

In Ref.~\cite{Ros15} a wide family of complex potentials that have the spectrum ${\cal E}_n = 2n+1$ of the quantum oscillator $V_{osc}(x)=x^2$ plus the energy eigenvalue $E_0=-1$ is reported (see Figure~\ref{oscc}). The energy eigenvectors $\vert \psi_k \rangle$ of this new potential form a bi-orthonormal basis of the corresponding space of states. In this communication we study the superposition properties of such a set of vectors. In particular, we construct a packet of $n+1$ adjacent states such that $\vert \psi_r \rangle$ is the eigenvector of lowest energy in the superposition. That is, the collection $\vert \psi_{r} \rangle, \vert \psi_{r+1} \rangle, \ldots, \vert \psi_{r+n} \rangle$ is used to tailor a pure state of the system that can be either classical (with nonnegative Wigner function) or nonclassical \cite{Ken04} in the limit of the quantum oscillator (i.e., when the imaginary part of the complex potential is turned off and the appropriate parameters are chosen). The Fourier coefficients $c_k$ associated with the states $\vert \psi_{r+k} \rangle$ in the superposition are such that $\vert c_k \vert^2$ is a binomial distribution delimited by $n$. We say that this bi-orthogonal superposition is an {\em optimized binomial state}. The concept of binomial states was introduced in \cite{Sto85} for the first $n+1$ eigenvectors of the quantum oscillator. The soundness of the approximation lies in the fact that such states represent light that is antibunched, sub-poissonian, and squeezed for certain parameter ranges \cite{Sto85} (see also \cite{Aha73}). Here we use the arbitrary collection of vectors described above and show that similar properties are obtained in the limit of the quantum oscillator, even if the generating set of vectors does not correspond to the first $n+1$ energy states. Moreover, we show that the summing of a very large number ($n\rightarrow\infty$) of elements in the collection leads to $r$-photon added coherent states.

In Section~\ref{spectrum} the main properties (bi-orthogonality, continuity equation, time-evolution) of the space of states $\vert \psi_k \rangle$  is revisited. Section~\ref{binomialS} deals with the tailoring of the classical and nonclassical states indicated above. Some concluding remarks are given in Section~\ref{concluding}.

\section{Complex oscillators with real spectrum}
\label{spectrum}

Let us consider the family of complex dimensionless potentials
\be
V_{\lambda}(x) =x^{2}-2-2\frac{d}{dx}\left[ \frac{b+2a \textnormal{Erf}(x)-i\sqrt{\pi}\lambda}{\sqrt{\pi}\alpha^{2}(x)} \right],
\label{pot1}
\ee
with $\lambda$ a real parameter and $\alpha(x)$ the real function
\be
\alpha(x) = e^{x^2/2} \left[ a \mbox{Erf}^2(x) + b \mbox{Erf}(x) + c \right]^{1/2}.
\label{pot2}
\ee
This family is reduced to the one reported by Mielnik \cite{Mie84} if $\lambda=0$ and the real parameters $a$, $b$ and $c$, are properly chosen \cite{Zel15}. For $\lambda \neq 0$, and any set of non-negative parameters $\{ a,b, c> \frac{b^2}{4a}\}$, the potential $V_{\lambda}(x)$ is exactly solvable \cite{Ros15} with spectrum $E_n = 2n-1$, $n=0,1,2,\ldots$ (see Figure~\ref{oscc}). Indeed, $V_{\lambda}(x)$ is a complex supersymmetric partner of the quantum  oscillator $V_{osc}(x)=x^2$, this last with spectrum ${\cal E}_n=2n+1$. The set of eigenfunctions is given by
\be
\psi_{n+1}(x)=\frac{1}{\sqrt{2(n+1)}}\left[ \frac{d}{dx} -\frac{\alpha'(x)}{\alpha(x)}+i\frac{\lambda}{\alpha^{2}(x)} \right]\varphi_{n}(x),  \quad  n\geq0,  
\label{sol1}
\ee
and
\be
\psi_{0}(x)=\frac{\kappa_0}{\alpha(x)} \exp \left[ i\lambda\int^{x}\alpha^{-2}(y) dy \right],
\label{sol2}
\ee
where $\kappa_0$ is a normalization constant and
\begin{equation}
\varphi_{n}(x)= \frac{e^{-x^{2}/2}}{\sqrt{2^{n}n!\sqrt{\pi}}} H_{n}(x)
\end{equation} 
are the wave-functions of the quantum oscillator. 

\subsection{Bi-orthogonality}

Although (\ref{sol1}) and (\ref{sol2}) do not form an orthonormal set, it is possible to construct a bi-orthogonal system $\{ \psi_n, \overline\psi_m \}_{n,m\geq 0}$ by considering that $H_{\lambda}=- \frac{d^2}{dx^2} + V_{\lambda}(x)$ and its Hermitian-conjugate $H_{\lambda}^{\dagger} \equiv \overline H_{\lambda}$ are formally different. The notion of bi-orthogonality arises when {\em orthogonality} requires two formally different sets of vectors. That is, if a given set $\{ \psi_n \}_{n\geq 0}$ cannot be orthogonal then one can consider the dual set $\{ \overline\psi_m \}_{m\geq 0}$ to get $(\overline \psi_m, \psi_n)=\delta_{m,n}$. As the wave-functions $\varphi_n(x)$ are real we get $\overline \psi_n(x)= \psi_n^*(x)$, with $z^*$ the complex-conjugate of $z \in \mathbb C$. Therefore, the following bi-orthonormal condition \cite{Ros15} holds
\be
\left( \psi^*_n, \psi_m \right) = \int_{\mathbb R} \psi_n(x) \psi_m(x) dx = \int_{\mathbb R} \overline \psi^*_n(x) \overline\psi^*_m(x) dx =\left( \overline\psi_n, \overline\psi^*_m \right) = \delta_{n,m}.
\label{orto}
\ee
Any arbitrary state $\phi(x)$ of the complex oscillator, as well as its dual $\overline \phi(x)$, can be expressed as a superposition of the fundamental solutions 
\be
\phi(x) = \sum_{k=0}^{\infty} c_k \psi_k(x), \quad \overline\phi(x) = \sum_{k=0}^{\infty} \overline{c}_k \overline\psi_k(x), \quad c_k, \overline{c}_k \in \mathbb C,
\label{super}
\ee
where the Fourier coefficients, $c_k$ and $\overline{c}_k$, are determined by the bi-products
\be
c_k= \left( \overline\psi_k, \phi \right), \quad \overline{c}_k = \left( \psi_k, \overline \phi \right).
\ee
In turn, the normalization condition 
\be
\left( \overline \phi, \phi \right)= \sum_{k=0}^{\infty} \overline{c}_k^* c_k=1
\ee
is fulfilled for the appropriate products $\overline{c}_k^* c_k \in \mathbb C$. We shall take $\overline{c}_k = c_k$ for simplicity. On the other hand, given $\phi(x)$ the conventional notions of probability density $\rho = \vert \phi \vert^2$ and probability current $J =i\left( \phi \frac{\partial \phi^*}{\partial x}- \phi^*\frac{\partial \phi}{\partial x} \right)$ lead to the continuity equation
\be
\frac{\partial J}{\partial x}+\frac{\partial \rho}{\partial t}=2\mbox{Im}(V_{\lambda}).
\ee
However, using the bi-orthogonal quantities $\rho_B = \overline\phi^* \phi$ and $J_B =i\left( \phi \frac{\partial \overline\phi^*}{\partial x}- \overline\phi^*\frac{\partial \psi}{\partial x} \right)$, one gets
\be
\frac{\partial J_B}{\partial x}+\frac{\partial \rho_B}{\partial t}=0.
\label{bicont}
\ee
If  $\lambda=0$ we have $\mbox{Im}(V_{\lambda}) =0$ and $\overline\psi_k = \psi_k$, so that $\rho_B$ and $J_B$ coincide with $\rho$ and $J$ respectively. From here, the limit to the quantum oscillator \cite{Zel15} gives $V_{\lambda=0} \rightarrow V_{osc}-2$, and $\psi_k \rightarrow \varphi_k$.

\subsection{Time-evolution}

The conventional time-evolution operator $U(t) = e^{-iH_{\lambda}t}$ is no longer unitary because $H_{\lambda}$ is not self-adjoint. To construct the dual of $U(t)$ in proper form, we use the bi-orthogonality introduced in the previous section. Namely, $\overline{U}(t) = e^{-i\overline{H}_{\lambda}t}$ is such that $\overline{U}^{\dagger}(t) U(t) =  I$ \cite{Zel15}. Then, the time-evolved vectors $\vert \phi(t) \rangle = U(t) \vert \phi(0) \rangle$ and $\vert \overline\phi (t) \rangle = \overline{U}(t) \vert \overline\phi (0) \rangle$ are such that
\be
\langle \overline\phi (t) \vert \phi(t) \rangle= \langle \overline\phi (0) \vert e^{i \overline{H}^{\dagger}_{\lambda} t} e^{-i H_{\lambda} t} \vert \phi (0) \rangle = \langle \overline\phi (0) \vert \phi(0) \rangle.
\ee
That is, the bi-norm of the initial state $\vert \phi(0) \rangle$ is invariant under time-evolution. Notice that this last result is compatible with the bi-orthogonal continuity equation (\ref{bicont}). In position-representation,  $\phi(x) = \langle x \vert \phi \rangle$ and $\overline\phi(x) = \langle x \vert \overline\phi \rangle$, one has
\be
\langle \overline\phi (t) \vert \phi(t) \rangle= \int_{\mathbb R} \overline\phi^*(x,t) \phi(x,t) dx = \int_{\mathbb R} \rho_B(x,t) dx=1.
\ee

\section{Optimized binomial states}
\label{binomialS}

Let us construct a superposition of $n+1$ adjacent states 
\be
\vert \phi_b \rangle = \sum_{k=0}^n c_k \vert \psi_{k+r} \rangle,
\label{finite}
\ee
where the non-negative integer $r \geq 0$ determines eigenvector $\vert \psi_r \rangle$ of the lowest energy that is included in the packet. If the Fourier coefficients $c_k$ are chosen such that
\be
\vert c_k \vert^2= \left( 
\begin{array}{c}
n\\ k
\end{array}
\right) p^{k}(1-p)^{n-k}, \quad 0\leq p \leq 1, \quad k\leq n, 
\label{finite2}
\ee 
then the state (\ref{finite}) is automatically bi-normalized $\langle \overline{\phi}_{b}\vert \phi_{b}\rangle=1$, with
\be
\left( 
\begin{array}{c}
n\\ k
\end{array}
\right) = \frac{\Gamma (n+1)}{\Gamma (k+1) \Gamma(n-k+1)}
\label{finite3}
\ee
the binomial coefficient. Therefore, the probability $\vert c_k \vert^2$ of finding the system in the state $\vert \psi_{k+r} \rangle$ is weighted by the probability $p$ of having success $k$ times in $n$ trials. We call the superposition (\ref{finite})--(\ref{finite3}) an {\em optimized binomial} state. The energy of such a wave-packet is characterized by the (bi-orthogonal) expectation value
\be
\langle H_{\lambda}\rangle_{b} \equiv \langle \overline{\phi}_{b}\vert H_{\lambda}\vert \phi_{b}\rangle=\sum_{k=0}^{n}\left( 
\begin{array}{c}
n\\ k
\end{array}
\right) p^{k}(1-p)^{n-k}(2k-1)=2np-1.
\ee
In turn, the time dependence of the superposition state is encoded in the Fourier coefficients
\begin{equation}
\phi_{b}(x,t)= \sum_{k=0}^{n} c_{k} (t) \psi_{k+r} (x), \quad c_k(t) = c_k e^{-2ikt},
\label{finite4}
\end{equation} 
where a global phase $e^{-i (2r-1)t}$ has been dropped. As we are going to show, the optimized binomial state (\ref{finite4}) is meaningful because any finite set of $n+1$ adjacent energy eigenvectors can be used to tailor either a classical-like or a quantum-like packet. 

We use the binomial distribution (\ref{finite2}) for two main reasons: (1) this permits the construction of wave-packets having a finite number of elements and (2) its limit for a large number of elements is well known. To be precise, in the limit $n \rightarrow \infty$ the expression (\ref{finite2}) goes to the normal (Gaussian) distribution if $p=\mbox{const}$, or to the Poisson distribution if $np=\mbox{const}$. The former case is avoided in our approach because it implies that the energy of the system increases as $n \rightarrow \infty$. In turn, the Poisson distribution  is reached by increasing the number of states but preserving the energy of the system. Thus, with $r=0$ and $n\rightarrow \infty$, the optimized binomial states (\ref{finite}) converge to the Glauber coherent states \cite{Gla07} in the quantum oscillator limit. For $r\neq 0$ and the same conditions the superposition (\ref{finite}) goes to an $r$-photon added coherent state. More details are given in Section~\ref{poissonS}.

\begin{figure}[htb]
\centering
\includegraphics[width=.9\textwidth]{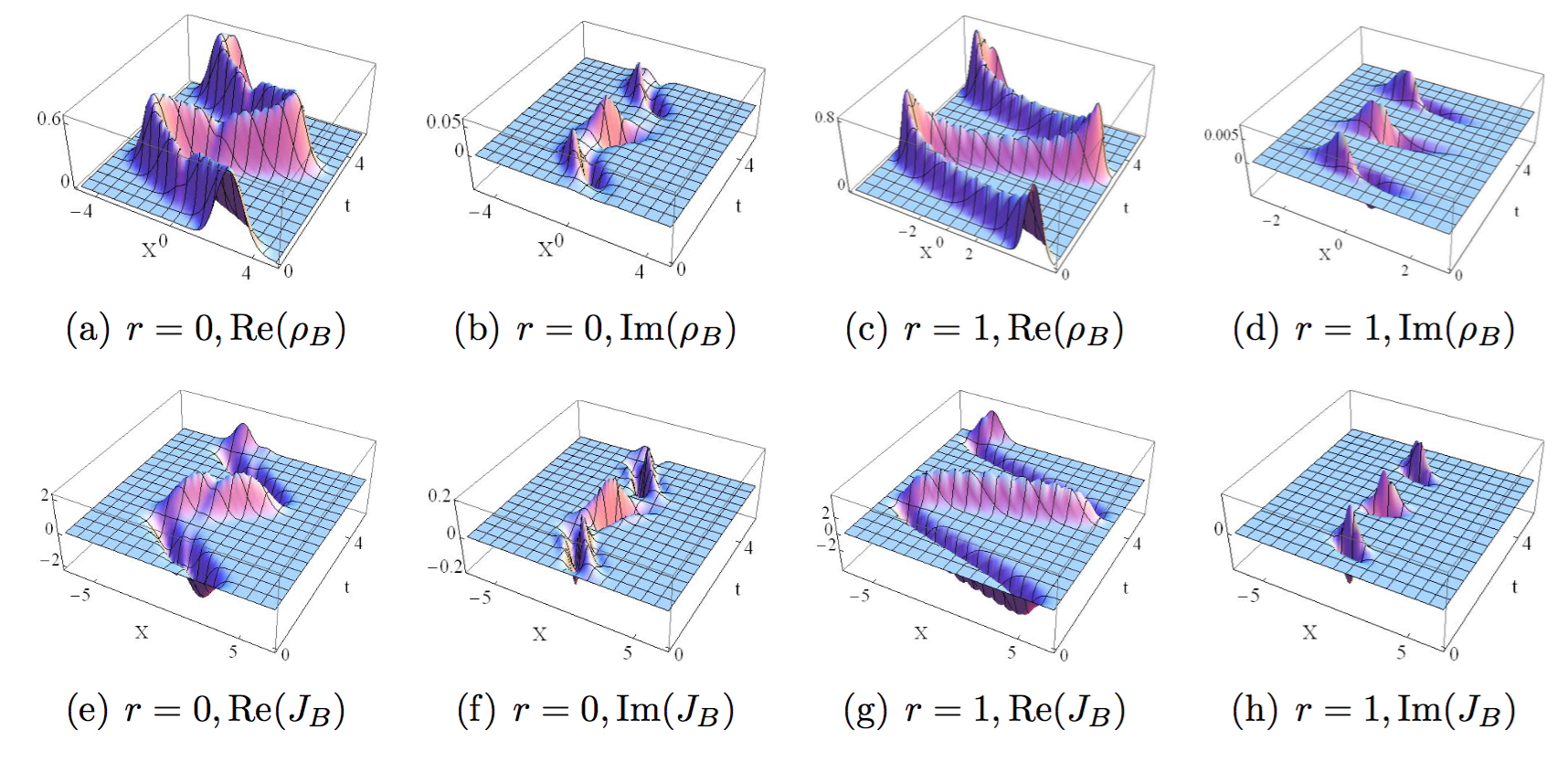}

\caption{\footnotesize 
Time-evolution of the bi-orthogonal probability density $\rho_B$ and current $J_B$ of the optimized binomial state (\ref{finite4}) for $n=30$ and the indicated values of $r$.  The first (last) two columns correspond to a packet of energy $\langle H_{\lambda} \rangle_b$ equal to 5 (29) and $p=0.1$ ($p=0.5$). }
\label{time1}
\end{figure}

The time-evolution of the related bi-orthogonal probability density $\rho_B$ and current $J_B$ is shown in Figure~\ref{time1} for $n=30$ and the indicated values of $p$ and $r$. The center of $\mbox{Re}(\rho_B)$ oscillates back and forth between two {\em turning points} while $\mbox{Im}(\rho_B)$ exhibits a zero (change of sign) that oscillates with a shorter amplitude. At a given time $t$, the integration of $\mbox{Im}(\rho_B)$ over all the real line is zero while the summing of $\mbox{Re}(\rho_B)$ gives 1. This last is because the time-variations of $\rho_B$ are compensated by the space-variations of the bi-orthogonal probability current $J_B$, as indicated in Eq.~(\ref{bicont}).

\begin{figure}[htb]
\centering
\includegraphics[width=.9\textwidth]{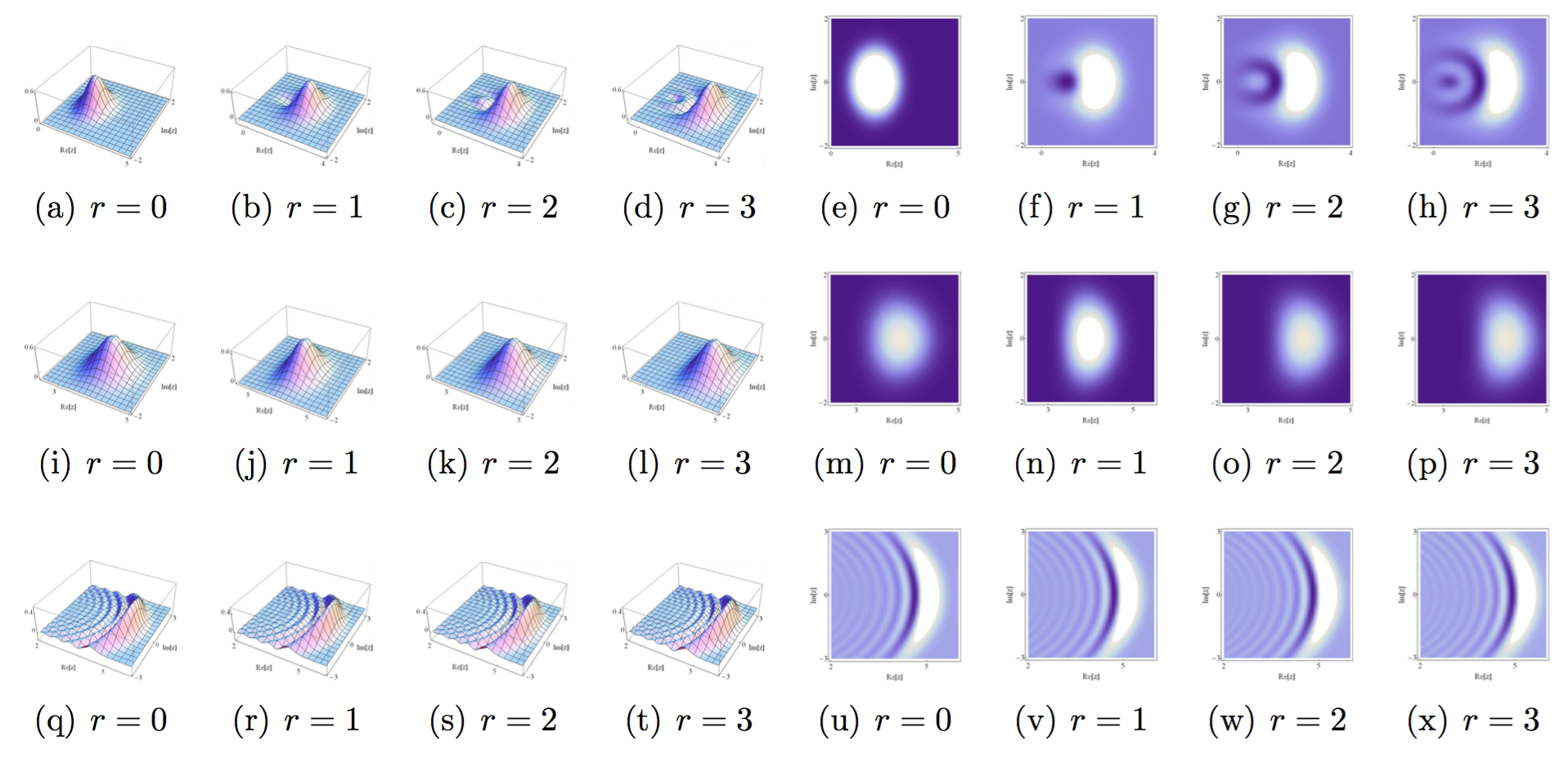}

\caption{\footnotesize 
Wigner distribution of the optimized binomial sates $\phi_b(x)$ in the limit of the quantum oscillator for $n=30$ and the indicated value of $r$. The last four columns are included as a reference and correspond to the density plot of the first four columns respectively. The probability of success $p$ has been chosen to be $p=0.1$ (first row), $p=0.5$ (second row) and $p=0.9$ (third row). }
\label{wigner1}
\end{figure}

In general, for $p<<1$ the major contribution to the packet $\rho_B$ is given by the state $\vert \psi_r \rangle$, so that we can construct a classical-like packet by taking $r=0$ (i.e., with a nonnegative Wigner function). Any other value of $r$, with $p<<1$, leads to a nonclassical-like packet. This last is illustrated in Figure~\ref{wigner1}, see first row, where we have plotted the related Wigner distribution \cite{Wig32} (see also \cite{Ken04}) in the limit of the quantum oscillator. For $p \approx 1$, see third row of Figure~\ref{wigner1}, we always have a quantum-like packet because the major contribution is given by the state $\vert \psi_{r+n} \rangle$. Quite interestingly, the situation changes for $p=0.5$ because the corresponding Wigner distribution is mainly non-negative, no matter the value of $r$, although this exhibits a squeezing that depends on $r$ (see second row of Figure~\ref{wigner1}).

\subsection{Poisson states}
\label{poissonS}

\begin{figure}[htb]
\centering
\includegraphics[width=.9\textwidth]{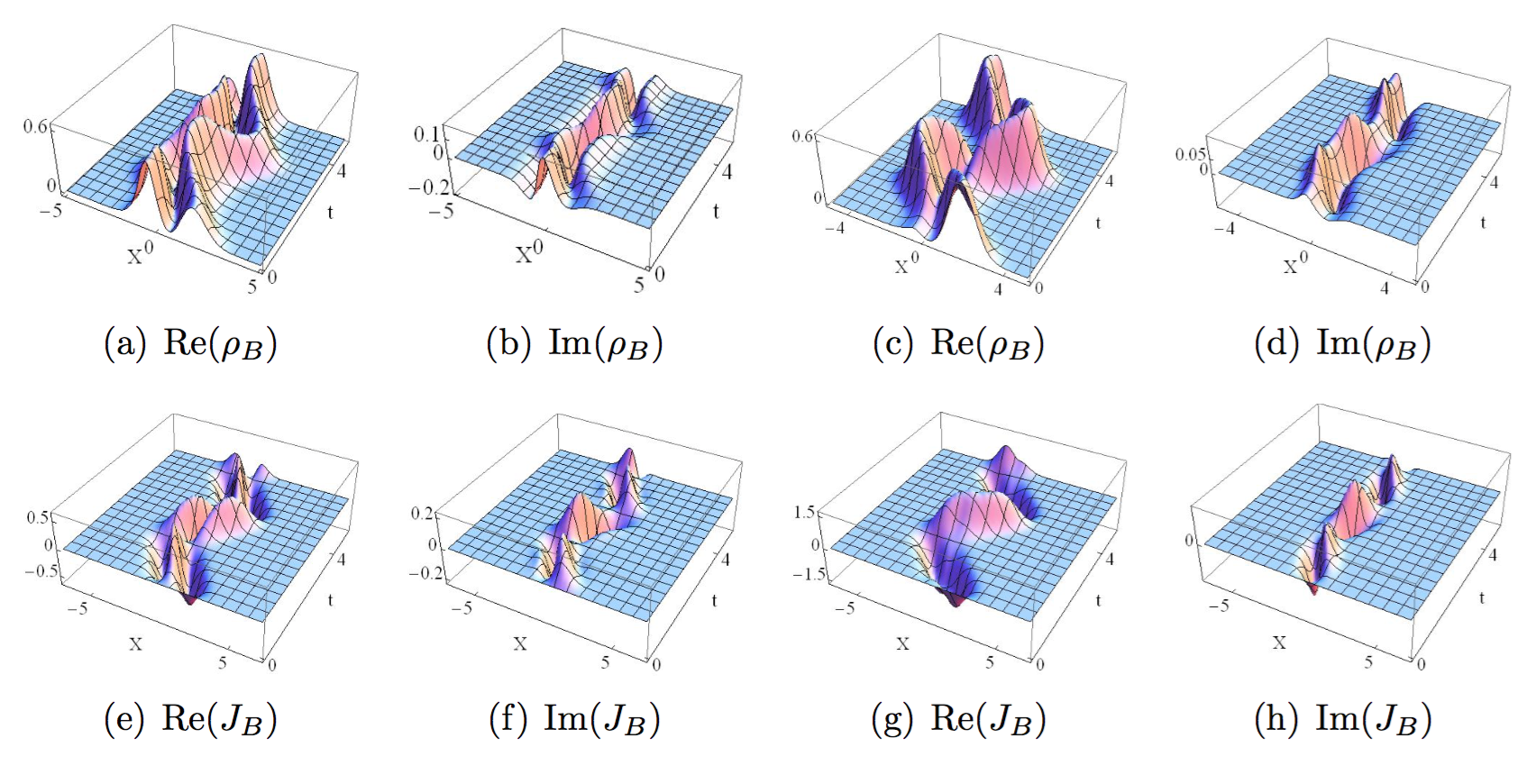}

\caption{\footnotesize 
Time-evolution of the bi-orthogonal probability density $\rho_B$ and current $J_B$ of a Poisson state $\phi_P$ with $\vert z \vert =1$. The first (last) two columns correspond to $r=0$ ($r=1$).
}
\label{time2}
\end{figure}

To get a superposition involving a large number of elements we use (\ref{super}) with
\begin{equation}
c_k=  \frac{e^{-\vert z \vert^{2}/2}}{\sqrt{k!}} \, z^k, \quad z\in\mathbb{C}.
\label{cs}
\end{equation}
That is, $\vert c_k \vert^2$ is the Poisson distribution with parameter $\vert z \vert^2$. As discussed above, such a distribution can be obtained from (\ref{finite2}) by considering $n\rightarrow \infty$ such that $np=\mbox{const}$. On the other hand, the Fourier coefficients (\ref{cs}) together with the Fock states $\vert \varphi_k \rangle$ of the quantum oscillator produce a Glauber coherent state with $\langle H_{osc} \rangle_z = 2 \vert z \vert^2+1$. We say that the superposition of states defined by (\ref{cs}) is a {\em Poisson state} $\phi_P$. The time-evolution of the related bi-orthogonal probability density and current is depicted in Figure~\ref{time2}. As in the previous case, the integration of $\mbox{Im}(\rho_B)$ over all the real line is equal to zero, and $\rho_B$ and $J_B$ satisfy the bi-orthogonal continuity equation (\ref{bicont}). Notice that in this case the real (imaginary) part of $\rho_B$ exhibits two local maxima (changes of sign) that evolve in time in oscillatory form.

\begin{figure}[htb]
\centering
\includegraphics[width=.9\textwidth]{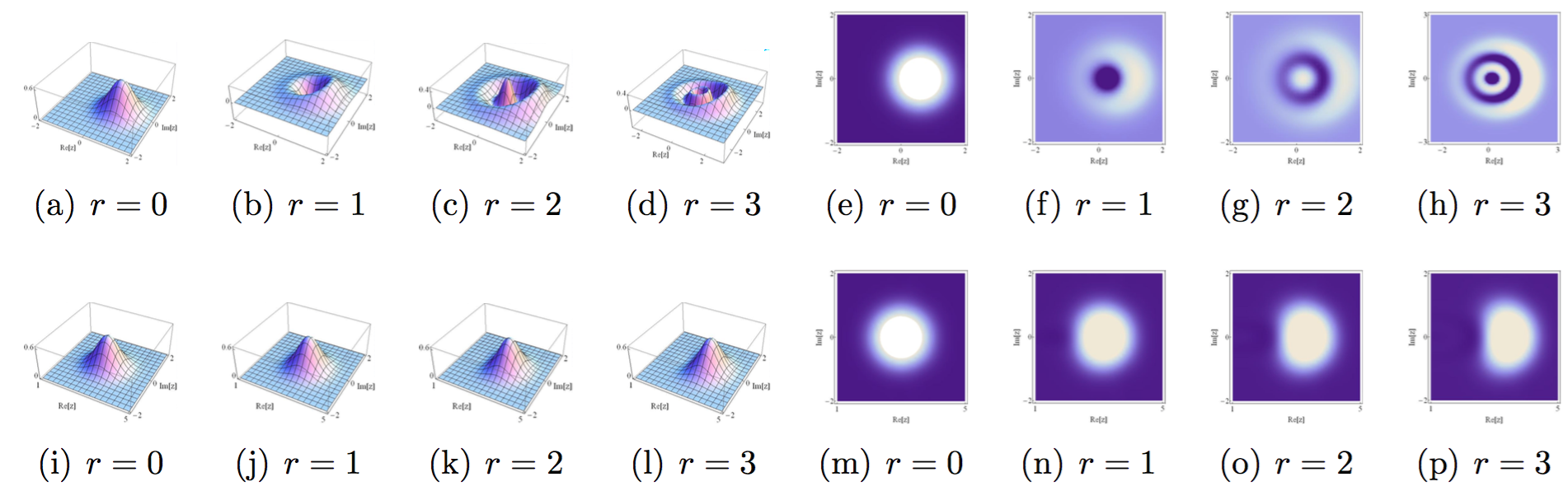}

\caption{\footnotesize 
Wigner distribution of the optimized Poisson sates $\phi_P(x)$ in the limit of the quantum oscillator for the indicated value of $r$. The last four columns are included as a reference and correspond to the density plot of the first four columns respectively. The parameter $z$ has been chosen real and equal to $0.6$ (first row) and $3$ (second row). }
\label{wigner2}
\end{figure}

In Figure~\ref{wigner2} we show the Wigner distribution of the Poisson states $\phi_P$ in the limit of the quantum oscillator. If $\vert z \vert =0$ the vector $\phi_P$ coincides with $\vert \psi_r \rangle$. For $0<\vert z \vert<1$, see first row of Figure~\ref{wigner2}, the sensitivity on the value of $r$ is noticeable. That is, the Poisson states $\phi_P$ behave as $r$-photon added coherent states \cite{Aga91}. The Wigner distribution is mainly positive for $\vert z \vert >>1$ and exhibits squeezing as $r$ increases.

\section{Concluding remarks}
\label{concluding}

The optimized binomial quantum states presented in the previous sections have been constructed as a superposition of $n+1$ (adjacent) energy eigenvectors of the complex oscillator (\ref{pot1}). Such a superposition has been tailored to represent either a classical or a nonclassical state in the limit of the quantum oscillator. Moreover, the superposition goes to an $r$-photon added coherent state by summing a very large number of energy eigenvectors ($n \rightarrow \infty$). On the other hand, the formal construction of the coherent states associated with the complex oscillator (\ref{pot1}) involves the algebraic properties of a fundamental set of operators that are not considered in this work, for more information see \cite{Zel15}. However, the results reported here are very close to the ones obtained by using either the Barut-Girardelo or the Perelomov approaches of coherent states \cite{Ros16}. The subject is still in progress, the details will be reported elsewhere. 

\subsection*{Acknowledgments}
The authors are grateful to the anonymous referee for valuable remarks. The financial support of CONACyT is acknowledged.


\end{document}